\documentclass[12pt]{iopart}
\usepackage{epsfig}

\def\la{\hbox{{\lower -2.5pt\hbox{$<$}}\hskip -8pt\raise
-2.5pt\hbox{$\sim$}}}
\def\ga{\hbox{{\lower -2.5pt\hbox{$>$}}\hskip -8pt\raise
-2.5pt\hbox{$\sim$}}}

\begin{document}

\title[Neutralino Annihilation at the Galactic Center Revisited]
{Neutralino Annihilation at the\\ 
Galactic Center Revisited}

\author{Roberto Aloisio$^a$, Pasquale Blasi$^{b,c}$ and
Angela V. Olinto$^{d,e}$}

\address{$^a$ INFN/Laboratori Nazionali del Gran Sasso,
SS 17bis, I-67010 Assergi (Italy)}

\address{$^b$ INAF/Osservatorio Astrofisico di Arcetri,
Largo E. Fermi 5, I-50125 Firenze (Italy)}

\address{$^c$ INFN/Sezione di Firenze, Italy}

\address{$^d$ Department of Astronomy \& Astrophysics,
The University of Chicago,\\ 
5640 S. Ellis Av., 60637 Chicago (USA)}

\address{$^e$ Center for Cosmological Physics, The University of Chicago}

\begin{abstract}
The annihilation of neutralino dark matter in the Galactic Center (GC) may 
result in radio signals that can be used to detect or constrain the dark 
matter halo density profile or dark matter particle properties. At the 
Galactic Center, the accretion flow onto the central Black Hole (BH) sustains 
strong magnetic fields that can induce synchrotron emission by electrons and 
positrons generated in neutralino annihilations during advection onto the BH. 
Here we reanalyze the radiative processes relevant for the neutralino 
annihilation signal at the GC, with realistic assumptions about the accretion 
flow and its magnetic properties. We find that neglecting these effects, 
as done in previous papers, leads to the incorrent electron and photon
spectra. We find that the magnetic fields associated with the
flow are significantly stronger than previously estimated. We derive the
appropriate equilibrium distribution of electrons and positron and the
resulting radiation, considering adiabatic compression in the accretion
flow, inverse Compton scattering off synchrotron photons (synchrotron
self-Compton scattering), and synchrotron self-absorption of the emitted
radiation. We derive the signal for a Navarro-Frenk-White (NFW) dark matter 
halo profile and a NFW profile with a dark matter spike due to the 
central BH. We find that the observed radio emission from the GC is 
inconsistent with the scenario in which a spiky distribution of neutralinos 
is present. We discuss several important differences between our calculations 
and those previously presented in the literature.
\end{abstract}

\section{Introduction}

The combination of cosmic microwave background anisotropy studies with
observations of distant type Ia supernovae and measurements of the large scale 
structure of the universe reveal that the energy density in our universe is
dominated  by dark energy ($\sim 70\%$) followed by a significant contribution
from matter ($\sim 27\%$). In addition, these studies show that the baryonic 
content of the universe is limited to less than $\sim 4 \%$ in agreement 
with big bang nucleosynthesis predictions. More specifically, recent WMAP 
results give the mean baryonic density in the universe to be within 
0.044 $\pm 0.004$ of the critical density while the matter density is 
0.27 $\pm 0.04$ of critical \cite{WMAP}. The remaining $\sim$ 0.23 of the 
matter density, i.e., the bulk of the matter in the universe, is believed 
to be a yet undiscovered form of dark matter.

Weakly interacting massive particles (WIMPs) are natural candidates for the
dark matter \cite{kamion}. Particles with masses around $\sim 100$ GeV that
interact only weakly have freeze-out densities in the required range. In 
addition, particle physics models that invoke supersymmetry generate a 
number of plausible WIMP candidates. In supersymmetric extensions of the 
standard model, the lightest supersymmetric particle may be stable due to 
conservation of R-parity enabling their survival to the present. In addition 
to massive and weakly interacting, dark matter particles are expected to be 
neutral. A class of neutral lightest supersymmetric particles is a combination
of gauginos and higgsinos, named the neutralino often represented by $\chi$.

A number of experiments are presently searching for the neutralino (see, 
e.g., \cite{bergst}). Accelerator  experiments have placed lower limits on 
the neutralino mass of $m_{\chi} \ga 37$ GeV  \cite{accel}.  The neutralino 
mass in the minimal supersymmetric extension of  the standard model that 
can explain the dark matter density is also bounded from above,  
$m_{\chi} \la 7$ TeV. In some constrained supersymmetric models 
$m_{\chi} \la 500$ GeV due to the  cosmological constraints from WMAP 
\cite{ellis}. 
Direct searches for neutralinos that cross the Earth are presently reaching 
the required sensitivity to probe the relevant range in  parameter space of 
supersymmetric models.

A good complement to accelerator and direct detection methods are indirect
searches based on the emission from neutralino self-annihilation in 
astrophysical environments 
\cite{bere1,bere2,berg1,gon1,calc,berg2,bert,ullio,blasi1,tasitol,cesar}.
Neutralinos self-annihilate at a rate proportional to the square
of the neutralino density. Thus, the highest density dark matter regions 
are the best candidates for indirect searches. In fact, the GC 
region may potentially be so dense that all neutralino models would be ruled 
out \cite{gon1}. This strong constraint arises in models where the 
super-massive black hole (BH) at the GC induces a strong dark 
matter density peak called {\it the spike}. The existence of such a spike 
is strongly dependent on the formation history of the Galactic Center BH 
\cite{merrit}. If the BH formed adiabatically a spike would be present, while 
a history of major mergers would most likely not allow the survival of a 
spike.

In contrast to the uncertain presence of a central spike in the dark matter
distribution, the central BH is known to induce an accretion flow of 
baryonic matter around its event horizon. The accretion flow carries magnetic 
fields, possibly amplified to near equipartition values due to the strong 
compression. The distribution of electrons and positrons (hereafter both 
called electrons) produced by neutralino annihilation at the GC 
would also be compressed toward the BH radiating through synchrotron and 
inverse Compton scattering off the local photon background.

By considering the injection of electrons, combined with radiative losses 
and adiabatic compression, we find the equilibrium spatial and spectral 
electron distribution and derive the expected radiation signal. We find that 
the synchrotron emission of electrons from neutralino annihilation 
ranges from radio and microwave energies up to the optical, in the
central more magnetized region of the accretion flow. At low frequencies,
synchrotron self-absorption (SSA) reduces the amount of  radiation
transmitted outwards. The resulting signal is stronger than the observed 
emission in the $10$ to $10^{5}$ GHz range for the case of a spiky dark 
matter profile while for a pure NFW profile the emission is below the 
observed level. 
Here, we do not attempt  to explore all 
possible spike profiles in the GC region, that might originate from the 
adiabatic compression of different types of {\it initial} dark matter 
profiles. Instead, we limit ourselves to the case of a spike arising 
from adiabatic
compression of a NFW profile in the gravitational field of the massive BH
at the GC (hereafter, {\it the spike case}.) in contrast to the pure NFW 
profile. 

Previous calculations of the type presented here were either incomplete 
or had unphysical values for some of the parameters. In particular, 
calculations presented in \cite{gon1} and \cite{bert} assume a magnetic 
field  that does not correspond to the equipartition field for  the accretion 
flow around the black hole. We show here that if this {\it incorrect} 
field is adopted, then the main processes responsible for reaching the 
equilibrium of relativistic electrons are synchrotron self-compton scattering 
(SSC) and adiabatic compression of the particles advected with the accretion 
flow, processes which were both ignored in previous calculations. On the other 
hand, if the {\it correct} magnetic field is adopted, the equilibrium 
distribution of the radiating electrons is determined by advection and 
synchrotron emission.

The paper is organized as follows: in \S \ref{sec:spatial} we describe the
spatial distribution of dark matter in the GC region. In 
\S \ref{sec:accret} we describe the accretion flow around the BH and its 
magnetic field. The injection of electrons and positrons from neutralino 
annihilation is considered in \S \ref{sec:inj}. In \S  \ref{sec:equi} we 
solve the transport equation for electron-positron pairs from neutralino 
annihilation in the presence of energy losses and adiabatic compression. 
Synchrotron self-absorption is discussed in \S \ref{sec:ssa}. Results of our 
calculations and comparison with observations are shown in 
\S \ref{sec:results} and comparison with previous work is made in 
\S \ref{sec:prev}. We conclude in \S \ref{sec:concl}.

\section{The spatial distribution of dark matter in the Galactic Center}
\label{sec:spatial}

The spatial distribution of dark matter in halos is still a matter of much
debate (see, e.g., \cite{tasit}). Numerical simulations suggest that
collisionless dark matter forms cuspy halos while some observations argue 
for a flat inner density profile \cite{saluc}. The standard numerical dark 
matter halo is the NFW profile \cite{nfw} which 
is expected to be universal. However, recent simulations have found more 
cuspy halos \cite{moore} as well as shallower profiles \cite{power}. At 
present, it is not clear that there is a universal dark matter halo profile, 
but the NFW case seems to represent well the range of plausible profiles.
Therefore, we assume that the NFW profile describes well the dark matter 
in our Galaxy. 

The NFW profile can be parametrized as
\begin{equation}
\rho(r)=\frac{\rho_0}{\frac{r}{R_c}\left[ 1+\frac{r}{R_c}\right]^2}.
\label{eq:nfw}
\end{equation}
where $R_c$ is the core radius and $\rho_0$ is a normalization constant. 
The two free parameters can be determined from the total mass of the Galaxy 
(that we take to be $2\times 10^{12} M_\odot$) and the velocity dispersion 
at the solar location which gives 
$\rho_{\odot}= \rho_{DM}(R_{\odot}) = 6.5 \times 10^{-25} {\rm g/cm}^3$ 
for $R_{\odot} = 8.5$ kpc.

There is now growing evidence for the presence of a supermassive black 
hole at the Galactic Center, with mass $\sim 2\times 10^6 M_\odot$. 
In fact most galaxies seem to harbor central black holes with comparable or 
even larger masses. The presence of a BH can steepen the density profile 
of dark matter by transforming the cusp at the GC into 
a spike of dark matter \cite{gon2}. The density profile of the spike region, 
where the gravitational potential is dominated by the BH is described by 
\begin{equation}
\rho'_{sp}(r)=\alpha_\delta^{\delta_{sp}-\delta}
\left( \frac{M}{\rho_\odot R_\odot^3}\right)^{(3-\delta)(\delta_{sp}-\delta)}
\rho_\odot g(r) \left(\frac{R_\odot}{r}\right)^{\gamma_{sp}}.
\label{eq:spike}
\end{equation}
Here, $\delta$ is the slope of the density profile of dark matter in the inner 
region ($\delta=1$ for a NFW profile), and 
$\delta_{sp}=(9-2\delta)/ (4-\delta)$. The coefficients $\alpha_\delta$ and 
$g(r)$ can be calculated numerically as explained in detail in \cite{gon2}. 
It is possible to identify a spike radius $R_{sp}$ where the spike density 
profile given in Eq. (\ref{eq:spike}) matches the NFW dark matter profile. In 
other words, at $R_{sp}$ the gravitational potential is no longer dominated 
by the central BH.

Neutralino annihilations affect the density profile in the spike by 
generating a flattening where the annihilation time becomes smaller than 
the age of the BH. This flattening occurs at the position
\begin{equation}
\rho_{core} = \frac{m_\chi}{\langle \sigma v \rangle_{ann}  t_{BH}} \ ,
\end{equation}
where  $\langle \sigma v\rangle_{ann}$ is the thermally averaged cross 
section  times velocity for neutralino annihilation. Cosmological arguments 
give  estimates of  $\langle \sigma v \rangle_{ann} \la 3 \times 
10^{-26} {\rm cm}^3 {\rm s}^{-1}$. The effect of annihilations on the spike 
density profile can be written as
\begin{equation}
\rho_{sp}(r) = \frac{\rho'_{sp}(r) \rho_{core}}{\rho'_{sp}(r) + \rho_{core}},
\end{equation}
which accounts for the flattening in the central region.

Several dynamical effects may weaken or destroy the spike in the GC 
\cite{kamion,merrit,zhao}, depending on the history of formation of the central
BH. If the spike is not formed or gets destroyed, the central region should 
be described by the cuspy profile as in the NFW case. Here we consider both
cases and show that the observed emission is stronger than the predictions 
for a NFW cusp, while the spike generates a signal well above the observed 
data.

\section{The accretion flow and its magnetic field}
\label{sec:accret}

We  model the accretion flow of gas onto the black hole at the center of our 
Galaxy following a simple approach described in \cite{melia}. More detailed
models of the accretion flow around the BH lead to corrections which are
negligible when compared to the uncertainties in the dark matter distribution.
In the model we adopt, the BH accretes its fuel from a nearby molecular cloud,
located at about $0.01$ pc from the BH. The accretion is assumed to be
spherically symmetric Bondi accretion with a rate of mass accretion of 
$\dot M=10^{22} \dot M_{22}~\rm g~s^{-1}$. The accretion onto the BH occurs 
with a velocity around the free-fall velocity, such that
\begin{equation}
v(r)=\sqrt{2 G M_{BH}/r} = c \left(\frac{R_g}{r}\right)^{1/2}
\end{equation}
where $R_g=2GM_{BH}/c^2=7.4\times 10^{11} 
(M_{BH}/2.5\times 10^6 M_\odot)~\rm cm$ is the gravitational radius of 
the BH and $M_{BH}$ is the BH mass. Therefore,
\begin{equation}
v(r)= 1.46 \times 10^8
\left(\frac{M_{BH}}{2.5\times 10^6 M_\odot}\right)^{1/2} 
\left(\frac{r}{0.01 \rm pc}\right)^{-1/2}~\rm cm~s^{-1}.
\end{equation}
Mass conservation then gives the following density profile:
\begin{equation}
\rho(r) = \frac{\dot M}{4 \pi r^2 v(r)} =  \frac{\dot M}{4 \pi R_g^2 c}
\left(\frac{r}{R_g}\right)^{-3/2} \ ,
\end{equation}  
such that
\begin{equation}
\rho(r) =  5.6\times 10^{-21} \dot M_{22}
\left(\frac{M_{BH}}{2.5\times 10^6 M_\odot}\right)^{-1/2}
\left(\frac{r}{0.01 \rm pc}\right)^{-3/2}~\rm g~cm^{-3} \ .
\end{equation}

Following \cite{melia}, we assume that the magnetic field in the accretion 
flow achieves its equipartition value with the kinetic pressure, namely
$\rho v^2/2 = B(r)^2/8\pi$. With this assumption,
\begin{equation}
B_{eq}(r)=\frac{\sqrt{\dot M c}}{R_g}
\left(\frac{r}{R_g}\right)^{-5/4}
=3.9\times 10^4 \dot M_{22} M_{BH}^{1/4} 
\left(\frac{r}{0.01\rm pc} \right)^{-5/4}~ \mu{\rm G} .
\label{eq:magnetic}
\end{equation}

It is believed that magnetic fields in the accretion flow will in general 
reach the equipartition values described in Eq. (\ref{eq:magnetic}). However, 
smaller fields may be reached if the equipartition is prevented somehow. 
In fact, previous authors \cite{gon1,bert} have assumed a lower field given 
by
\begin{equation}
B_{low}(r)=1 \mu {\rm G} \left(\frac{r}{1 {\rm pc}}\right)^{-5/4} = 
3 \times 10^2 ~\mu{\rm G}~\left(\frac{r}{0.01 {\rm pc}}\right)^{-5/4}
\simeq 10^{-2} \, B_{eq} ~,
\label{eq:lowB}
\end{equation}
although this field was actually referred to as the 
{\it equipartition field}. In what follows, we assume the equipartition 
field as calculated in Eq. (\ref{eq:magnetic}), and derive the signal from 
neutralino annihilation. In \S \ref{sec:prev}, we address the case of 
assuming a lower filed such as in Eq. (\ref{eq:lowB}).

\section{Neutralino annihilations: the injection of electrons and positrons}
\label{sec:inj}

Neutralino annihilations can result in many different final states (see, e.g. 
\cite{ullio} and references therein). Given the large uncertainties in the
dark matter distribution, we chose to make the simplifying assumption that
the annihilation channels are dominated by quarks and antiquarks, which in 
turn generate mainly  pions:
\begin{equation}
\chi + \chi \to q + \bar q \to \rm pions.
\end{equation}
Approximately $1/3$ of the pions are neutral and promptly decay to gamma 
rays. About $2/3$ of the pions are charged and result in the production of 
electrons, positrons and neutrinos. Although much attention has been devoted 
in the past literature to the gamma rays from $\pi^0$ decays, electrons and 
positrons can result in copious production of nonthermal radiation in the 
presence of magnetic fields and a photon background.

We simplify the annihilation treatment further by assuming that the spectrum 
of charged pions in a single neutralino annihilation is described by the 
Hill \cite{hill} quark fragmentation:
\begin{equation}
W_\pi (E_\pi) = \frac{5}{4}\frac{1}{m_\chi} x^{-3/2} (1-x)^2\ ,
\label{eq:frag}
\end{equation}
where $x=E_\pi/m_\chi$. In Eq. (\ref{eq:frag}), a $2/3$ factor accounts for 
the fraction of charged pions compared to neutral ($1/3$) in each neutralino
annihilation event, and a factor $2$ accounts for two jets, each with energy
$m_\chi$, the neutralino mass. With these simplifying assumptions, the 
spectrum of generated electrons appears to possess the main features found 
with a full treatment of the neutralino annihilations (see \cite{ullio} for 
a description of the full calculation), although the latter should be used 
for a more quantitative description of the particle physics aspects of the 
problem.

The spectrum of electrons (and positrons) from the $\pi^\pm$ decays is 
calculated by convoluting the spectrum of pions and muons. For relativistic 
electrons the electron spectrum reads
 
\begin{eqnarray}
W_e(E_e)=\int_{{\rm max}(E_e,m_{mu})}^{m_\chi} dE_\mu
\int_{E_{\pi}^{{\rm min}}(E_\mu)}^{E_{\pi}^{{\rm max}}(E_\mu)}
d E_\pi W_\pi(E_\pi)\cdot \nonumber \\
\qquad\qquad\qquad \cdot \frac{m_\pi^2}{m_\pi^2-m_\mu^2}
\frac{1}{\sqrt{E_\pi^2-m_\pi^2}} \frac{dn_e(E_e,E_\mu,E_\pi)}{dE_e}
\end{eqnarray}
where, neglecting the muon polarization, we get 
\begin{equation}
\frac{dn_e(E_e,E_\mu,E_\pi)}{dE_e}=\frac{1}{E_\mu\beta}
\left\{ \begin{array}{ll}
2\left [\frac{5}{6}-\frac{3}{2}\epsilon^2+\frac{2}{3}\epsilon^3\right ]
&{\rm if} ~~~\frac{1-\beta}{1+\beta}\le \epsilon \le 1 \\
\frac{4\epsilon^2\beta}{(1-\beta)^2}
\left [3-\frac{2}{3}\epsilon\left(\frac{3+\beta^2}{1-\beta}\right)^2\right ]
&{\rm if} ~~~0\le\epsilon\le\frac{1-\beta}{1+\beta},
\end{array}
\right.
\end{equation}
with $\epsilon=\frac{2}{1+\beta}\frac{E_e}{E_\mu}$. Here $\beta$ is the pion
speed and $E_e$ and $E_\mu$ are the total energies of electrons and muons
respectively. The two limits of integration 
$E_{\pi}^{{\rm min}}(E_\mu)$ and 
$E_{\pi}^{{\rm max}}(E_\mu)$ can be derived by inverting the following 
equations:
\begin{equation}
E_\mu\le \frac{E_\pi}{2m_\pi^2}
\left [m_\pi^2 (1+\beta)+m_\mu^2 (1-\beta)\right ]
\end{equation}
\begin{equation}
E_\mu\ge \frac{E_\pi}{2m_\pi^2}
\left [m_\pi^2 (1-\beta)+m_\mu^2 (1+\beta)\right ]~.
\end{equation}

Finally, the injection of new electrons at the distance $r$ from the BH at 
energy $E$ can be written as follows:
\begin{equation}
Q(E,r)=\frac{1}{2} \left (\frac{\rho_{DM}(r)}{m_{\chi}} \right)^2 
W_e(E) \langle \sigma v\rangle_{ann},
\end{equation}
where the density of dark matter has the profile $\rho_{DM}(r)$ discussed
in \S \ref{sec:spatial}, with or without a spike.  In this equation, the 
factor $1/2$ is introduced to avoid double counting of the annihilating 
neutralinos, as pointed out in \cite{cesar}.

\section{The equilibrium distribution of electrons near the BH}
\label{sec:equi}

The spectrum of particles at a position $r$ in the accretion flow around 
the BH is the result of the injection of newly produced electrons at the same 
position, radiative losses of these electrons, and the adiabatic compression 
that may enhance their momentum as they move inward. Here we neglect spatial 
diffusion, which occurs on larger time scales.

The transport equation including all these effects can be written as 
follows:
\begin{equation}
v(r)\frac{\partial f}{\partial r} - \frac{1}{3 r^2}\frac{\partial}{\partial r}
\left[r^2 v(r)\right]~p~ \frac{\partial f}{\partial p}+
\frac{1}{p^2}\frac{\partial}{\partial p}\left[ p^2 \dot p(r,p) f\right]
=Q(r,p),
\label{eq:transport}
\end{equation}
where $f(r,p)$ is the equilibrium distribution function of electrons injected
according to $Q(r,p)$ that lose energy radiatively as described by the
function $\dot p(r,p)= dp(r,p)/dt$. Here $v(r)=-c(r/R_g)^{-1/2}$ is the inflow 
velocity. The equation can be solved analytically if the electrons remain 
relativistic everywhere in the fluid, which is a good approximation for the 
range of frequencies of the radiation that we are interested in.

\begin{figure}
\begin{center}
\includegraphics[width=0.7\textwidth]{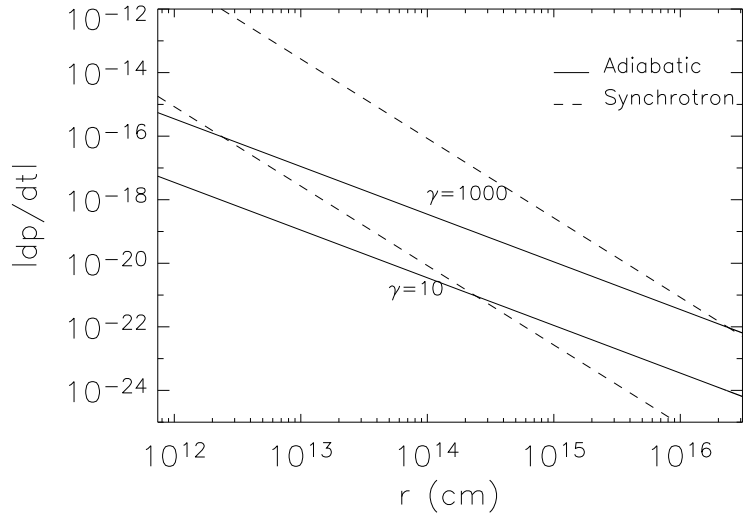}
\caption{Comparison between the time scales for adiabatic compression (solid
lines) and synchrotron losses (dashed lines) for $\gamma=10$ and
$\gamma=1000$.}
\label{fig:losses}
\end{center}
\end{figure}

In Eq. (\ref{eq:transport}), the term
\begin{equation}
\dot p_{ad} = -\frac{1}{3} p \nabla v(r) =
-\frac{1}{3 r^2} p \frac{\partial}{\partial r}\left[r^2~v(r)\right]
\end{equation}
describes the rate of change of momentum of a particle at the position $r$
due to adiabatic compression in the accretion flow. The rate of adiabatic
momentum enhancement should  be compared with the rate of losses due to
synchrotron emission:
\begin{equation}
\dot p_{syn}(r,p)=\frac{4}{3}\sigma_T \frac{B^2(r)}{8\pi} \gamma^2,
\label{eq:syn}
\end{equation}
where $\sigma_T$ is the Thomson cross section and $\gamma$ is the Lorentz
factor of the electron. The magnetic field $B_{eq}(r)$ depends on $r$ as
described in Eq. (\ref{eq:magnetic}).
Comparing the absolute values of the adiabatic compression rate and the
rate of synchrotron losses, we find that the effect of adiabatic
compression dominates at
$$\frac{r}{R_g}> 32.7 ~ \gamma \ . $$
For this estimate we used the fiducial value of the BH mass ($2.5
\times 10^6 M_\odot$) and the accretion rate ($\dot M=10^{22} \rm g~s^{-1}$).
A graphical comparison of the time scales for adiabatic compression (solid
lines) and synchrotron losses (dashed lines) is plotted in Fig.
\ref{fig:losses} for $\gamma=10$ and $\gamma=1000$.

It is clear from Fig. \ref{fig:losses} that for Lorentz factors $\gamma<1000$,
there is a region (large radii) for which the particles injected at $r$
are advected inward faster than they lose energy through synchrotron emission.
If instead of using $B_{eq}(r)$ from Eq. (\ref{eq:magnetic}),
we adopted $B_{low}(r)$ from Eq. (\ref{eq:lowB}) as in \cite{gon1,bert},
the relation for the dominance of the adiabatic compression would become
$$\frac{r}{R_g} > 2.3\times 10^{-3}~\gamma \ .$$
In this case, the adiabatic compression (ignored in previous calculations)
becomes more important than synchrotron losses at most energies. The final
spectrum would then differ substantially from that obtained in 
\cite{gon1,bert}, as we discuss in \S \ref{sec:prev}.

In order to solve the transport equation, Eq. (\ref{eq:transport}), we discuss
different loss processes using the equipartition field $B_{eq}$. We first
consider synchrotron losses ($\dot p_{syn}$), followed by inverse Compton 
losses ($\dot p_{ICS}$) and synchrotron self-compton scattering 
($\dot p_{SSC}$).

The rate of synchrotron losses from Eq. (\ref{eq:syn}) is given by:
$$
\dot p_{syn}(r,p) ~ c = 1.6 \times 10^{-18} \left(\frac{r}{0.01\rm pc}
\right)^{-5/2} ~ \gamma^2 ~ \rm erg ~ s^{-1}  \ .
$$
In the Thomson regime, losses due to Inverse Compton Scattering (ICS) off a
photon background with energy density $U_{ph}$ has the following form
\begin{equation}
\dot p_{ICS}(r,p)=\frac{4}{3}\sigma_T U_{ph}(r) \gamma^2 \ .
\end{equation}
ICS  dominates synchrotron losses only if  $B^2 >  8\pi U_{ph}$.

Assuming that $U_{ph}$ is independent of $r$, i.e., a fixed photon background, 
ICS becomes important at large radii ($\sim 0.01 \rm pc$) and only if
$U_{ph}\ga 10^4 \rm eV ~ cm^{-3}$. Such strong photon background is unlikely to
be present at the GC region. For comparison, the CMB radiation has
$U_{CMB}\approx 0.25 \rm eV ~ cm^{-3}$ while the optical background has
$U_{opt}\approx 1 \rm eV ~ cm^{-3}$. If ICS off a fixed background is not
dominant at large radii, it becomes even less important at small radii when
compared to synchrotron losses. Consequently, we safely neglect the role of 
ICS off photons of fixed photon backgrounds.

The electrons, radiating in the strong magnetic field near the BH generate a
photon background that can become quite intense. The rate of losses due to 
ICS of electrons off the photons generated through synchrotron emission by 
the same electrons is
\begin{equation}
\dot p_{SSC}(r,p)=\frac{4}{3}\sigma_T U_{ph}^{syn}(r) \gamma^2,
\label{eq:SSC}
\end{equation}
where the photon energy density generally depends on the radius $r$. The 
photon density $U_{ph}^{syn}(r)$ is a nonlinear function of the distribution 
$f(r,p)$. In other words, the term $\dot p$ in Eq. (\ref{eq:transport}) depends
on $f(r,p)$ when synchrotron self-Compton scattering is included. If this
contribution cannot be neglected, an analytical solution of the transport
equation becomes unattainable.

Given the distribution function $f(r,p)$, one can calculate the synchrotron
emissivity  $j(\nu,r)$ (energy per unit volume, per unit frequency, per unit
time). The photon energy density at the position $r$ is then proportional to 
the integration over all lines of sight of the emissivity, with the possible
synchrotron self-absorption taken into account at each frequency. However,
the distribution function $f(r,p)$ is not known {\it a priori}, and the 
problem becomes intrinsically nonlinear. The approach that we follow here 
is to start with neglecting  synchrotron self-Compton scattering and check 
{\it a posteriori} whether the assumption is correct in the situation at 
hand.

An analytic solution of the transport equation can be derived when 
$\dot p(r,p)$ is dominated by synchrotron losses as in Eq. (\ref{eq:syn}).
In this case, the equation admits the following analytical solution:
\begin{eqnarray}
f(r,p) = \frac{1}{c}\left(\frac{r}{R_g}\right)^{-2}
\int_r^{R_{acc}} d R_{inj} 
\left(\frac{R_{inj}}{R_g}\right)^{5/2}\cdot \nonumber\\
\qquad\qquad\qquad \cdot \left(\frac{p_{inj}[p,r,R_{inj}]}{p}\right)^{4}
Q(R_{inj},p_{inj}[p,r,R_{inj}]).
\end{eqnarray}

We rewrite the synchrotron losses in the form:
$$\dot p_{syn}=k_0~\left(\frac{r}{R_g}\right)^{-5/2}~p^2\qquad \ ,$$
where $k_0= \sigma_T B_0^2/6\pi (mc)^2$.
The function $p_{inj}[p,r,R_{inj}]$ corresponds to the injection
momentum of an electron injected at the position $R_{inj}$ that arrives 
at the position $r$ with momentum $p$. This injection momentum can be 
obtained by inverting, with respect to $p_{inj}$ the solution of the equation 
of motion of the electron, in the presence of adiabatic compression and 
radiative losses:
\begin{equation}
\frac{dp}{dr} = \frac{k_0}{c}\left(\frac{r}{R_g}\right)^{-2} p^2
-\frac{1}{2 R_g}~p~\left(\frac{r}{R_g}\right)^{-1}.
\end{equation}
The solution of this equation, with initial condition
$p[r=R_{inj},p_{inj},R_{inj}]=p_{inj}$ is
\begin{equation}
p[r,p_{inj},R_{inj}] = p_{inj} \left[
\frac{2 k_0}{3 c} \frac{R_g^2}{r}p_{inj} \left[1-\left(\frac{r}{R_{inj}}
\right)^{3/2}\right] +\left (\frac{r}{R_{inj}}\right )^{1/2}\right ]^{-1}~.
\label{eq:dif_sol}
\end{equation}
In the absence of synchrotron energy losses (when $k_0 = 0$), particle momenta 
only change due to adiabatic compression, and the momentum of a particle 
changes according with the well known $p=p_{inj}(r/R_{inj})^{-1/2}$, valid 
for the case of free fall.

The joint effect of the energy gain due to the adiabatic compression and 
the energy losses due to synchrotron emission generates a new energy scale 
in the system. Introducing
\begin{equation}
p_m=\frac{3 r c}{2 k_0 R_g^2}=\frac{9\pi r (m c)^2}{\sigma_T \dot{M}}~,
\label{eq:pm}
\end{equation}
we can rewrite $p[r,p_{inj},R_{inj}]$ as:
\begin{equation}
p[r,p_{inj},R_{inj}] =  p_{inj}
\left[\frac{p_{inj}}{p_m}\left(1-\left(\frac{r}{R_{inj}}\right)^{3/2}\right )
+ \left (\frac{r}{R_{inj}}\right )^{1/2}\right]^{-1} \ .
\end{equation}

From this expression it is clear that, at any fixed position $r$,
adiabatic compression dominates over synchrotron losses if the injection
momentum is appreciably lower than $p_m$. 
In this case, the electron energy increases
while the electron moves inward, until the rate of synchrotron losses
becomes important. The opposite happens when the electrons are injected
at momenta larger than $p_m$, since synchrotron losses are important
from the time of injection.

The momentum $p_m$ can be interpreted as the momentum where the two competing 
effects of adiabatic heating and synchrotron losses balance each other. Thus,
particles accumulate at momentum $p_m$. This phenomenon depends on the 
distance from the GC: at large distances from the BH the 
momentum $p_m$, which scales linearly with radius, is large and the local 
rate of injected electrons is low, therefore, the accumulation is small. At 
small distances the accumulations at $p_m$ becomes more evident.

\begin{figure}
\begin{center}
\begin{tabular}{ll}
\includegraphics[width=0.5\textwidth]{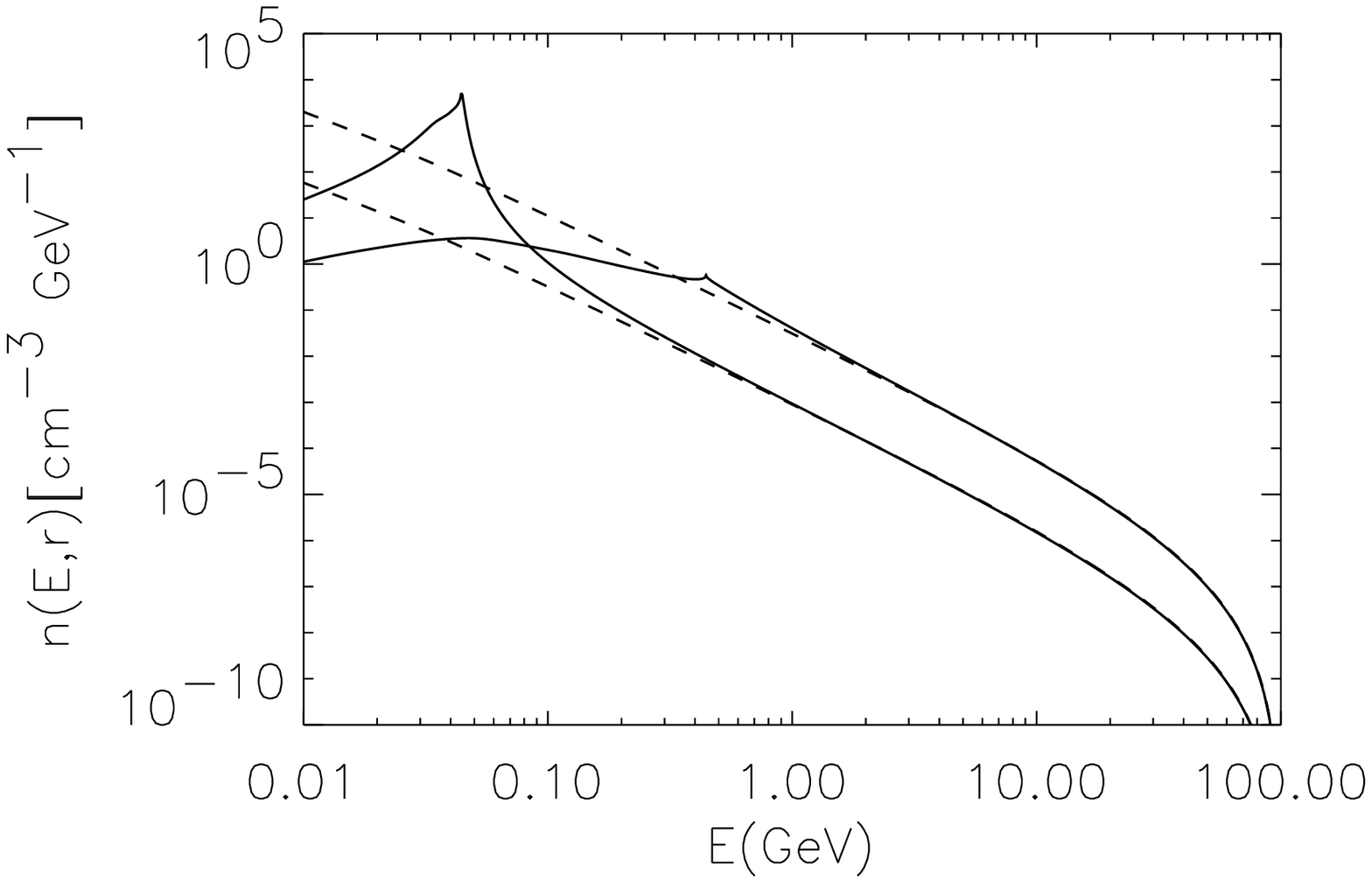}
&
\includegraphics[width=0.5\textwidth]{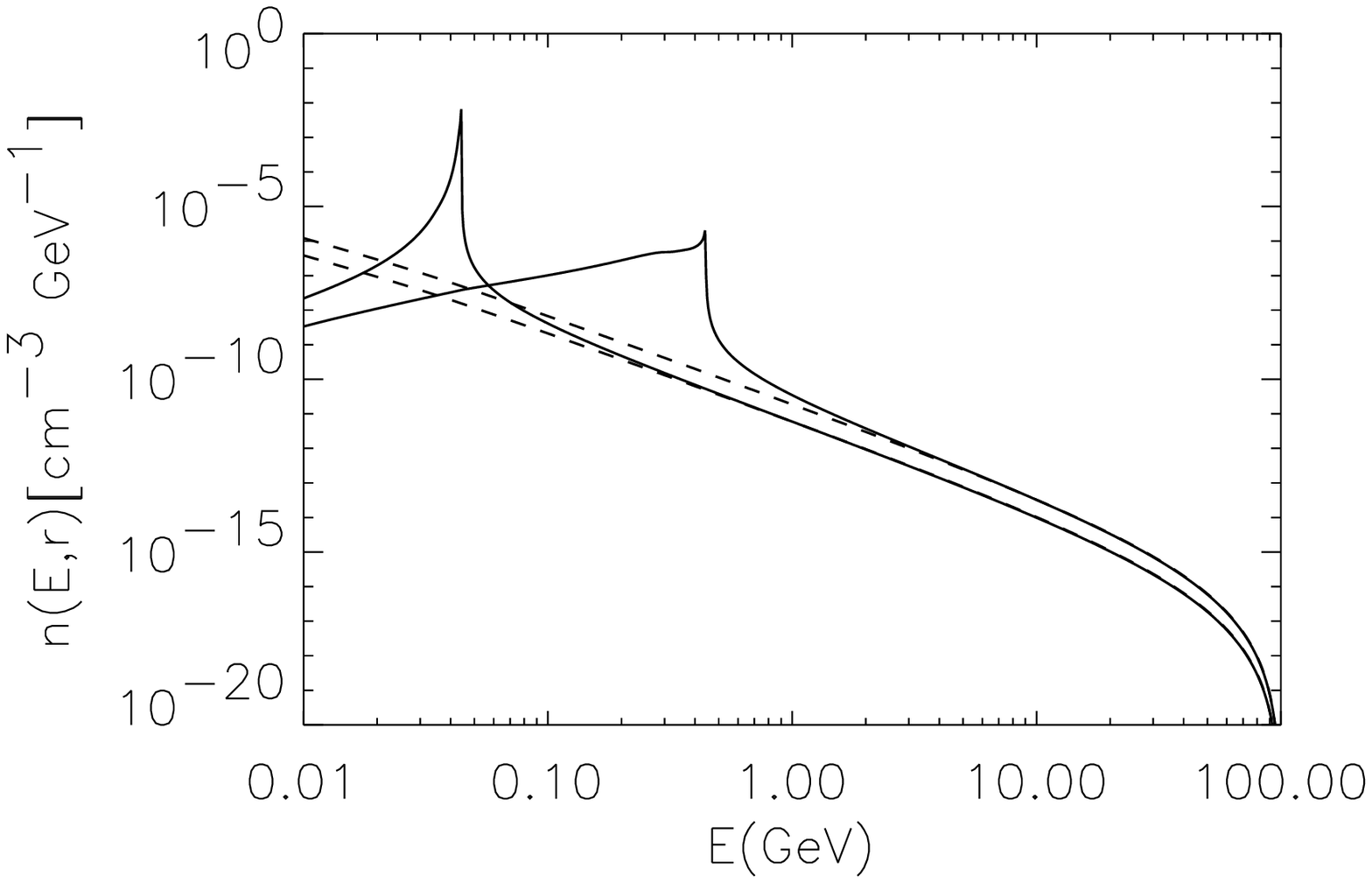}
\\
\end{tabular}
\caption{Electron density per unit energy as a function of energy
(solid lines) at $r=10^{3}R_g$ (upper curve) and $r=10^{4}R_g$
(lower curve). Superimposed (dashed curves), we plot the function
$n(E,r)$ obtained without the effect of advection. The left panel
illustrates the spike case, while the right panel applies to the
NFW case.}
\label{fig:nEr}
\end{center}
\end{figure}

We define now the equilibrium electron spectrum $n(E,r)$ at the position
$r$, which is related to $f(E,r)$ through the relation
$$n(E,r)dE=4\pi p^2 f(p,r)dp.$$
In Fig. \ref{fig:nEr} we plot $n(E,r)$ as a function of the electron energy
at two different radii, $r=10^{3}R_g$ (upper solid curve) and $r=10^{4}R_g$
(lower solid curve). The dashed lines illustrate the solution of the 
transport equation when adiabatic compression is switched off and only
synchrotron losses are included (as in \cite{bert}). The dark matter density
profile is assumed to be characterized by a spike at the center \cite{gon2}
in the left panel of Fig. \ref{fig:nEr}, and by an NFW density profile
\cite{nfw} in the right panel of Fig. \ref{fig:nEr}. The neutralino mass 
assumed in Fig. \ref{fig:nEr} was 100 GeV, while the annihilation 
cross-section was chosen to be $10^{-27}\rm cm^3 s^{-1}$. 
The accumulation effect described above manifests itself through the 
appearance of the spiky feature at the momentum $p_m$. The accumulation 
is less pronounced at large radii, as expected.

We conclude this section by addressing the issue of the synchrotron self-
Compton scattering. As explained above, this effect cannot be accounted for
in an analytical approach to the transport equation, since it is intrinsically
nonlinear. Instead, we check a posteriori if neglecting SSC was a good 
assumption. Given the symmetry of the BH region, the photon energy density as 
a function of $r$ can be written as:
\begin{equation}
U_{ph}^{syn}(r)=\frac{1}{c}\left [
\frac{1}{r^2}\int_{r_{min}}^r dr' r'^2 \int d\nu j(\nu,r')
+\int_r^{\infty}dr' \int d\nu j(\nu,r') \right ]~,
\label{eq:Uph}
\end{equation}
where $j(\nu,r)$ is the synchrotron emissivity.

\begin{figure}
\begin{center}
\includegraphics[width=0.7\textwidth]{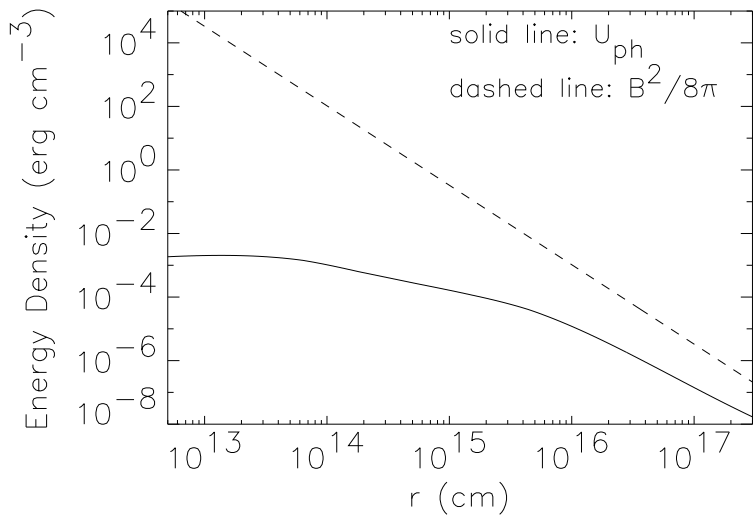}
\caption{Comparison between the energy density of the synchrotron emitted
photons $U_{ph}$ (solid line) and the energy density associated to the 
magnetic field $B^2/8\pi$ (dashed line).}
\label{fig:Uph}
\end{center}
\end{figure}

This energy density can now be compared with the magnetic energy density 
at the same location, $B^2(r)/8\pi$. The results are plotted in 
Fig. \ref{fig:Uph}: the solid line represents the photon energy density 
$U_{ph}^{syn}(r)$, while the magnetic energy density is plotted as a dashed 
line. The calculations are carried out for a dark matter density profile with 
the spike at the center and a neutralino mass of 100 GeV. The magnetic energy 
always dominates over the photon energy, although at large distances from the 
BH the separation between the two curves reduces to about one order of 
magnitude. The curves in Fig. \ref{fig:Uph} are obtained without taking into 
account the synchrotron self-absorption (see next section) at low frequencies,
therefore the calculated photon energy density should be considered as an 
upper limit. On the basis of this argument, we can conclude that neglecting 
SSC was a reasonable approximation for the scenario at hand.

\section{Synchrotron Self-Absorption}
\label{sec:ssa}

Synchrotron radiation can be reabsorbed by the radiating electrons when the
system is sufficiently compact. This phenomenon of synchrotron 
self-absorption (SSA) is particularly effective at low frequencies. The 
radiation transfer across the GC region can be described by  
\cite{ryby}:
\begin{equation}
\frac{d I (\nu,s)}{ds}=-\chi(\nu,s) I(\nu,s) + \frac{1}{4\pi} j(\nu,s)~,
\label{eq:SSA1}
\end{equation}
where $I(\nu,s)$ is the radiation intensity at frequency $\nu$ along a fixed
line of sight, $s$ is the linear coordinate along the line of sight,
$\chi(\nu,s)$ is the absorption coefficient and $j(\nu,s)$ is the emissivity.

For sufficiently high frequencies, where the absorption coefficient
is small, the radiation intensity is simply
\begin{equation}
I(\nu,s)=\frac{1}{4\pi}\int ds j(\nu,s)~.
\label{eq:noSSA}
\end{equation}

The absorption coefficient can be written in the form:
\begin{equation}
\chi(\nu,s)=-\frac{c^2}{8\pi\nu^2} \int dE P(\nu,E) E^2
\frac{\partial}{\partial E} \left [\frac{n(E,r(s))}{E^2} \right ]
\label{eq:chi1}
\end{equation}
where $P(\nu,E)$ is the synchrotron power radiated per unit frequency by one
electron of energy $E$, and $n(E,r(s))$ is the electron density per unit 
energy discussed in the previous section, evaluated at the position $r(s)$.
Integrating equation (\ref{eq:chi1}) by parts, using the vanishing electron
density at $E=m_\chi$ we obtain:
\begin{equation}
\chi({\nu},s)=\frac{c^2}{8\pi\nu^2}\int dE \frac{n(E,r)}{E^2}
\frac{\partial}{\partial E} \left [E^2 P(\nu,E)\right ]~,
\label{eq:chi2}
\end{equation}
where
\begin{equation}
P(\nu,E)=\frac{\sqrt{3} e^3 B(r)}{mc^2} \left (\frac{\nu}{\nu_c(E)}\right )
\int_{\left (\frac{\nu}{\nu_c(E)}\right )}^{\infty} dy K_{5/3}(y) ~.
\end{equation}
Here $K_{5/3}$ is the modified Bessel function of order $5/3$ and
\begin{equation}
\nu_c(E)=\left (\frac{3 e B(r)}{4\pi m c}\right )
\left (\frac{E}{mc^2}\right )^2 ~.
\label{eq:nu_c}
\end{equation}
The absorption coefficient is therefore
\begin{equation}
\chi(\nu,s)=\frac{c^2}{4\pi}\frac{\sqrt{3} e^3 B(r)}{mc^2}
\int dE\frac{n(E,r)}{E\nu_c^2(E)} K_{5/3}\left (\frac{\nu}{\nu_c(E)} \right )~.
\end{equation}

\begin{figure}
\begin{center}
\includegraphics[width=0.7\textwidth]{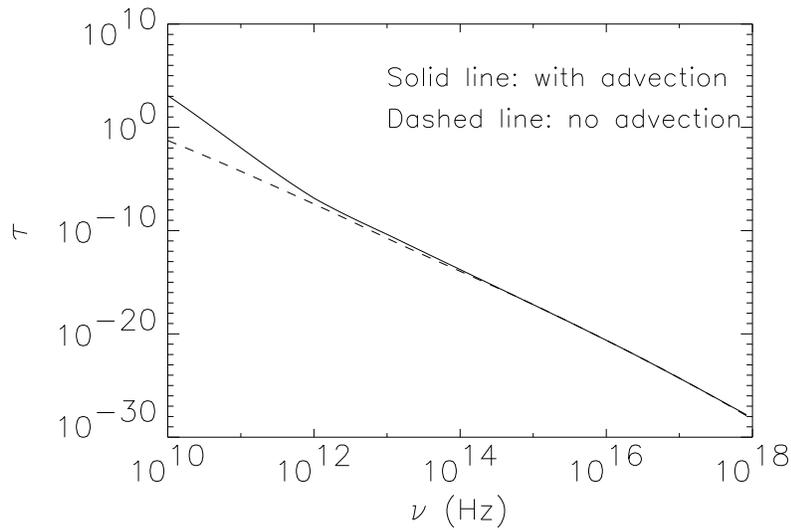}
\caption{Optical depth as a function of the photon frequency in two cases:
1) advection plus synchrotron losses (solid line); 2) synchrotron losses only
(dashed line). The calculation was carried out for a line of sight tangent
to the event horizon of the black hole.}
\label{fig:tau}
\end{center}
\end{figure}

The effect of the SSA can be assessed by evaluating the opacity to the 
absorption process at fixed frequency $\nu$, defined as
\begin{equation}
\tau_\nu(b)=\int ds \,  \chi[\nu, r(b,s)],
\end{equation}
where $b$ is the distance of the line of sight from the line crossing the GC.

Clearly SSA becomes important when $\tau_\nu(b)$ approaches or exceeds unity.
When $\tau_\nu(b)\ll 1$, SSA can be neglected and the radiation intensity 
can be simply evaluated as in Eq. (\ref{eq:noSSA}). Whether the SSA is 
important or not depends also on the line of sight considered: in general the 
GC can become opaque to SSA at different frequencies for 
different lines of sight. The observed radiation is then the result of a 
line of sight integration of the radiation intensity, as described below.

In Fig. \ref{fig:tau} we plot the opacity $\tau_{\nu}$ for a line of sight
tangent to the event horizon of the BH (solid line), namely with $b=R_g$.
In this figure, we chose the dark matter density with  a spike, 
$m_{\chi}=100$ GeV and $\langle \sigma v \rangle_{ann}=10^{-27}$ cm$^3$/s.
The dashed line in Fig. \ref{fig:tau} represents the opacity to SSA if only
synchrotron losses are included while adiabatic compression is neglected 
(as in \cite{bert}). For the correct calculation in which both synchrotron 
losses and advective compression are included, the SSA becomes important for 
frequencies $10^{10}-10^{11}$ Hz. When only synchrotron losses are considered,
SSA becomes important only at much lower frequencies. 

The dependence of the SSA on neutralino parameters scales with
$n(E,r) \propto \langle \sigma v \rangle_{ann} ~ m_{\chi}^{-3/2}$. At fixed
$\langle \sigma v \rangle_{ann}$, the electron density drops when $m_{\chi}$
increases, and SSA decreases as well. 
Finally, in the case of an NFW density profile without a spike, the rate 
of electron injection remains such that the effect of SSA can always be 
neglected at the frequencies of interest.

\section{Results}
\label{sec:results}

\begin{figure}
\begin{center}
\begin{tabular}{ll}
\includegraphics[width=0.5\textwidth]{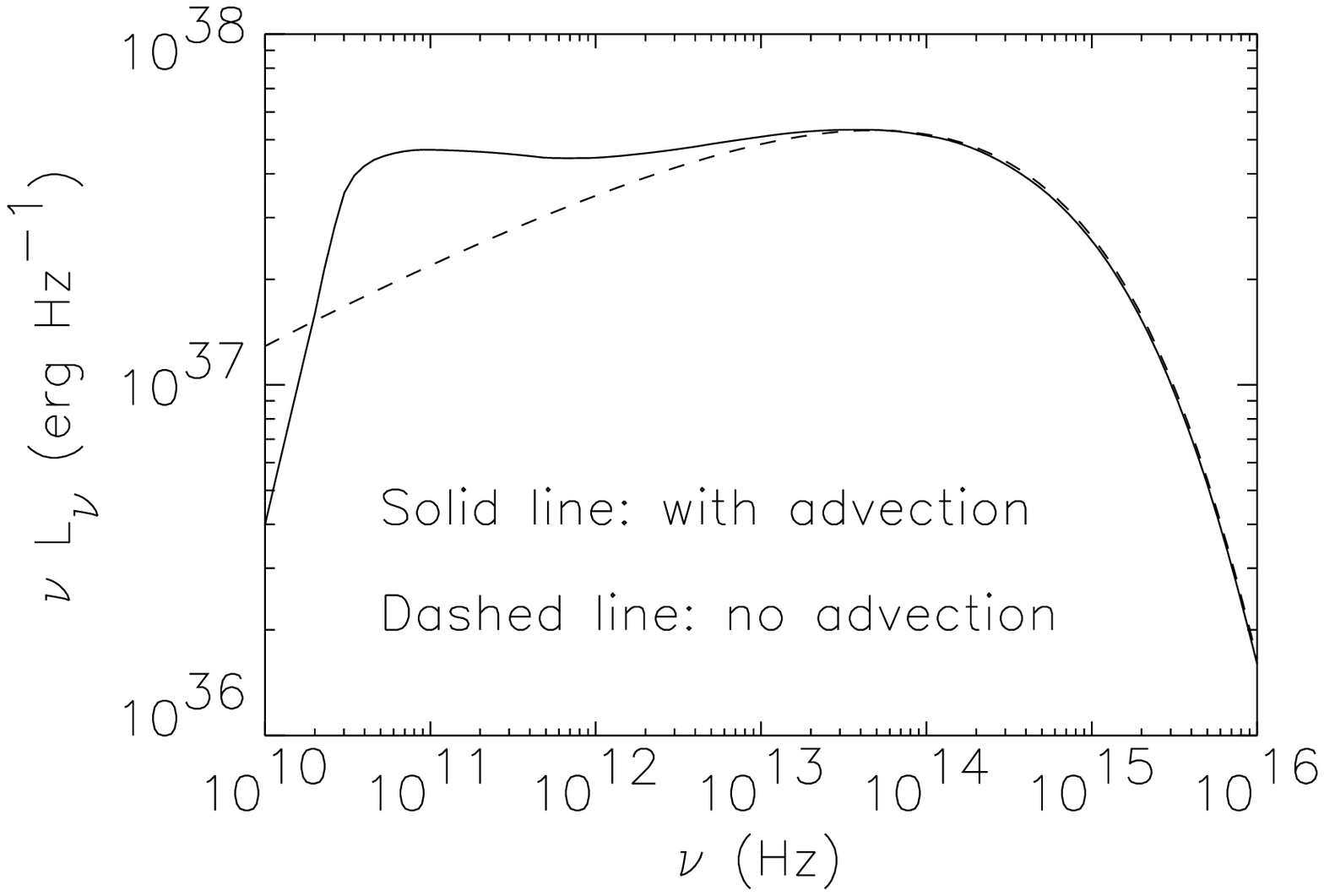}
&
\includegraphics[width=0.5\textwidth]{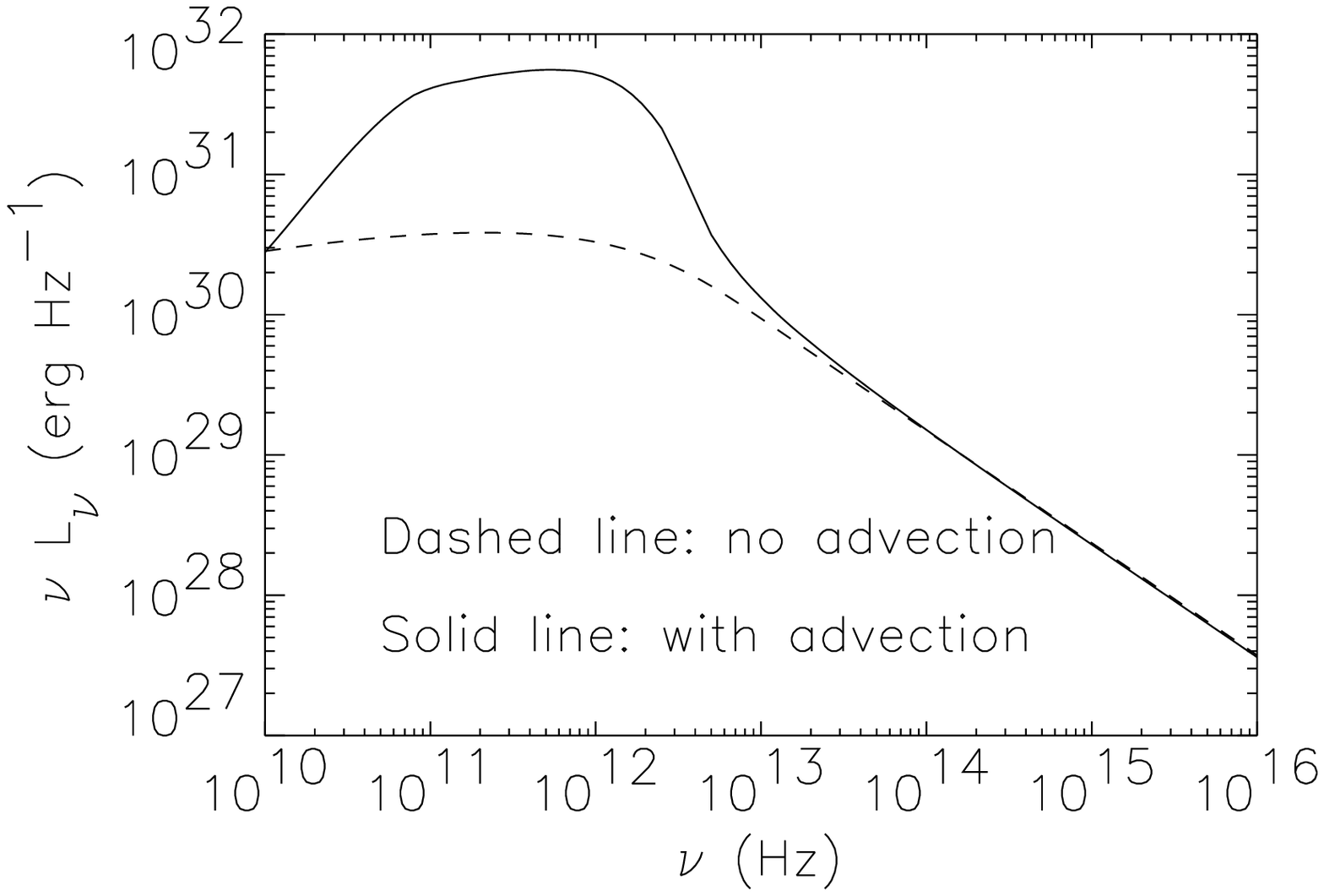}
\\
\end{tabular}
\caption{Emitted luminosity in two cases: 1) advection and synchrotron losses
(solid line); 2) only synchrotron losses (dashed line). The computation is
performed with $m_{\chi}=100$ GeV and $\langle \sigma v \rangle_{ann}=10^{-27}$
cm$^3$/s. The left panel represents the Spike case while right panel the 
NFW case.}
\label{fig:Lnu1}
\end{center}
\end{figure}

In this section, we present our results in terms of the synchrotron luminosity 
produced by relativistic electrons from neutralino annihilations in the GC. We 
considered both cases of a spiky density profile and a NFW density profile. 
In order to compute the radiation intensity $I_{\nu}$, we solved 
Eq. (\ref{eq:SSA1}) in the case of the spiky
density profile in which, as discussed in the previous section, the SSA effect
is relevant. In the case of the NFW density profile we simply used
equation (\ref{eq:noSSA}). Integrating the radiation intensity $I_{\nu}$ over
the sky (namely, over all lines of sight to the observer), one obtains the
luminosity:
\begin{equation}
L_{\nu}=4\pi \int db 2\pi bdb I_\nu(b)~.
\label{eq:Lnu1}
\end{equation}
In Fig. \ref{fig:Lnu1}, we fix  $m_{\chi}=100$ GeV and 
$\langle \sigma v \rangle_{ann}=10^{-27}$ cm$^3$/s, and show the luminosity
for the spiky case (left panel) and the NFW density profile (right panel).
The continuous line shows the case in which advection is taken into account,
while the dashed line shows the case in which only the synchrotron energy
losses are accounted for.

A comparison of the dashed and continuous curves in Fig. \ref{fig:Lnu1} shows
that the combined effect of advection and synchrotron losses appears at 
frequencies up to $10^{14}$ Hz. In this frequency range, the electron density 
is much higher when advection is included as compared to the pure synchrotron 
case. This effect produces an enhancement of the emitted luminosity of about 
one order of magnitude. In addition, we can see that SSA decreases the emitted
radiation in the frequency range $10^{10} - 10^{11}$ Hz. This effect becomes 
relevant only when advection is taken into account.

\begin{figure}
\begin{center}
\begin{tabular}{ll}
\includegraphics[width=0.5\textwidth]{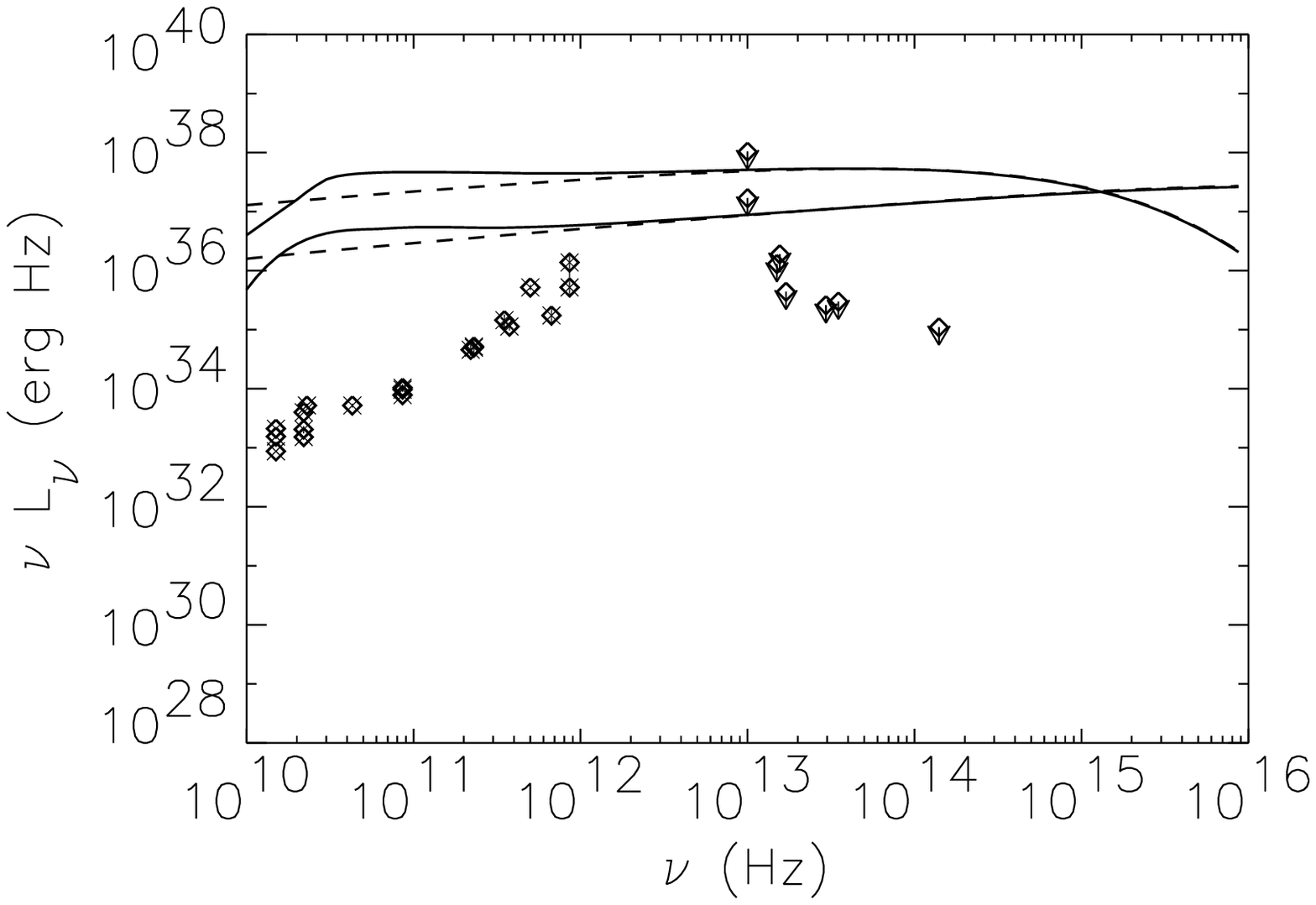}
&
\includegraphics[width=0.5\textwidth]{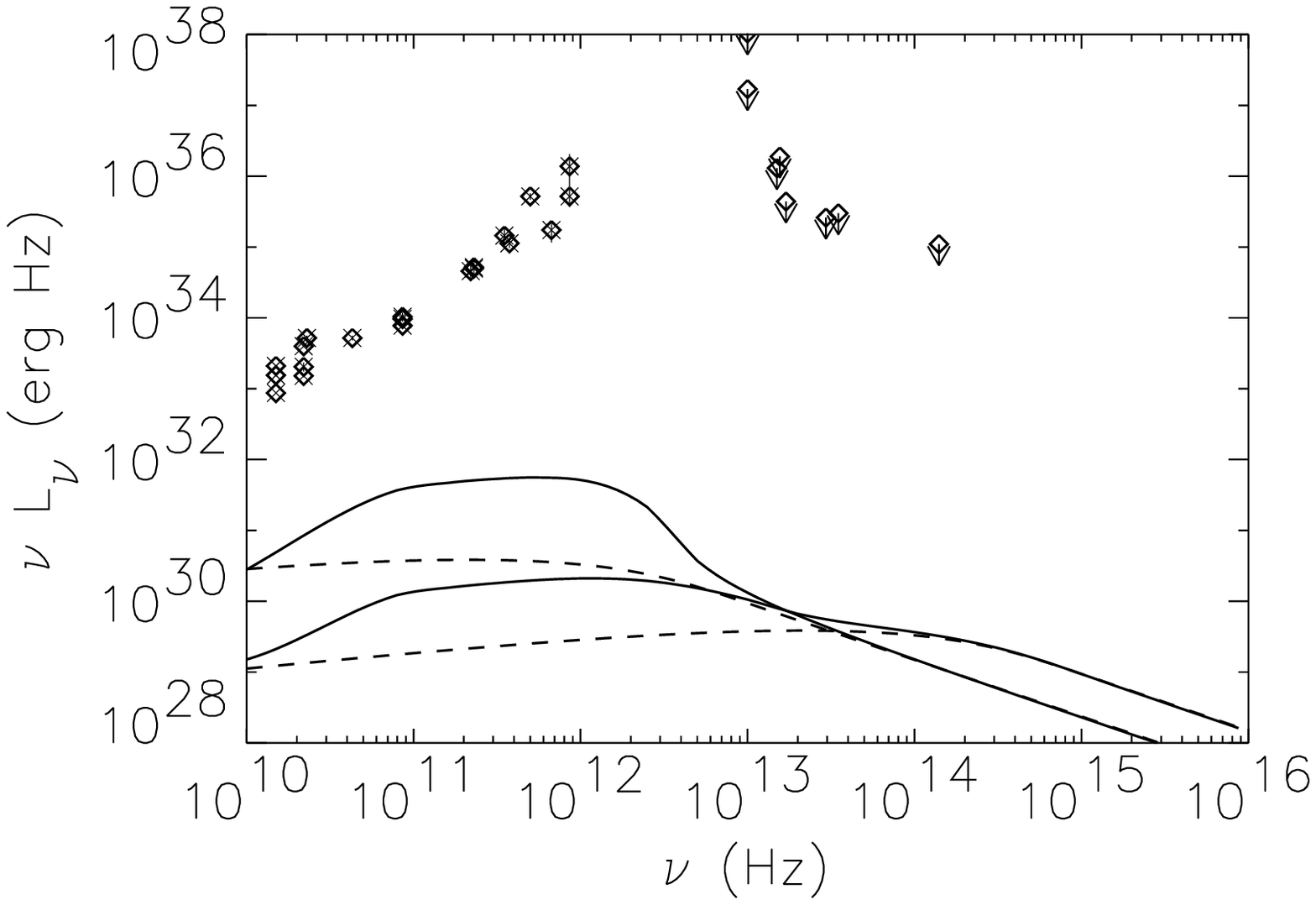}
\\
\end{tabular}
\caption{Luminosity, compared with experimental data from Sgr A$^*$,
in the case of the spiky density profile (left panel) and NFW density
profile (right panel). Dashed curves represent the luminosity obtained
neglecting the effect of advection. The computation is performed with
$m_{\chi}=100$ GeV (upper curves) and $m_{\chi}=1$ TeV (lower curves).
The annihilation cross section is
$\langle \sigma v \rangle_{ann}=10^{-27}$ cm$^3$/s.}
\label{fig:Lnu2}
\end{center}
\end{figure}

In order to constrain the shape of the dark matter density profile at the GC 
or the parameters of the dark matter particle, we compare the calculated
luminosities with the observations. The well-studied source at the GC, 
Sgr A$^*$, has been observed in a large range of frequencies, from the radio 
to the near-infrared. Data from $10^9$ Hz to $10^{14}$ Hz \cite{data} are 
displayed in Fig. \ref{fig:Lnu2}. In the same figure we show the two cases of
spiky (left panel) and NFW (right panel) density profiles, for
$m_{\chi}=100$ GeV (upper curves) and $m_{\chi}=1$ TeV
(lower curves) and $\langle \sigma v \rangle_{ann}=10^{-27}$ cm$^3$/s.
In Fig. \ref{fig:Lnu2} we also show the results for the cases in which only
synchrotron losses are taken into account (dashed curves).

From Fig. \ref{fig:Lnu2}, it is clear that a spike in the GC induces much
stronger emission than the observed flux. Therefore, either there is no
spike in the dark matter profile or neutralinos are not the dark matter.
This conclusion qualitatively confirms what found in previous studies 
\cite{gon1}, although the magnetic field and radiative transfer here are 
significantly different from what was assumed in \cite{gon1},  which 
results in numerically different conclusions (see next section for a more 
detailed discussion of the comparison of our results with those obtained 
in previous literature).

The parameter space for neutralinos ($m_{\chi},\langle \sigma v \rangle_{ann}$)
has been studied extensively \cite{bergst,para}. From accelerator and 
cosmological constraints $m_{\chi} \in$ [37 GeV, 7 TeV] and 
$\langle \sigma v \rangle_{ann} \la 3 \times 10^{-26}  {\rm cm}^3{\rm /s}$. 
Since the luminosity scales with the neutralino mass and the annihilation 
cross section through the electron density, and 
$n(E,r)\propto \langle \sigma v \rangle_{ann} ~ m_{\chi}^{-3/2}$, the 
luminosity we calculated can be lowered by more than two orders of magnitude 
when $m_{\chi}$ is increased from 100 GeV to 7 TeV. The luminosity increases 
linearly for a fixed $m_{\chi}$ as $\langle \sigma v \rangle_{ann}$ increases.
A factor of 30 increase is within the allowed parameters as
$\langle \sigma v \rangle_{ann}$  is changed from $10^{-27}$ cm$^3$/s to 
$3 \times 10^{-26}$ cm$^3$/s. Exploring the parameter space can change our 
results from the fiducial choices (100 GeV, $10^{-27}$ cm$^3$/s) by two 
orders of magnitude in either direction.

The large enhancement of the neutralino density due to the spiky profile
produces a synchrotron luminosity that is difficult to reconcile with
observations. Moreover, this conclusion only gets stronger when advection is
included. The SSA effect does not affect the discrepancy between a neutralino
spike and the observations, since it changes the luminosity only at 
frequencies in the range $10^{10} - 10^{11}$ Hz.

The situation is completely different for the case of the NFW density profile.
In this case, the synchrotron luminosity is always less than the experimental
data as can be seen in Fig. \ref{fig:Lnu2}. NFW is consistent with observations
even if $\langle \sigma v \rangle_{ann}= 3 \times 10^{-26}$ cm$^3$/s is 
considered.

In this paper, we only considered two possibilities for the dark matter density
profile: a less concentrated hypothesis (NFW) and an extremely concentrated 
central region (spike). As we stressed above, the spike we  considered 
results from the adiabatic compression of a NFW dark matter
profile in the gravitational field of the central BH. Less extreme spikes 
can be 
generated if the dark matter is initially distributed in a shallower manner
compared with the NFW profile. On the other hand,  there are cuspier original 
profiles, such as the Moore et al.  profile (which scales as $r^{-3/2}$). A 
simple Moore profile should generate a central luminosity in between the ones 
illustrated in this paper, while a central BH in such a profile will result in 
more pronounced spike. In some of these alternative cases, 
observations are likely to place more stringent limits on the neutralino 
parameters.

\section{Comparison with previous results}
\label{sec:prev}

\begin{figure}
\begin{center}
\includegraphics[width=0.7\textwidth]{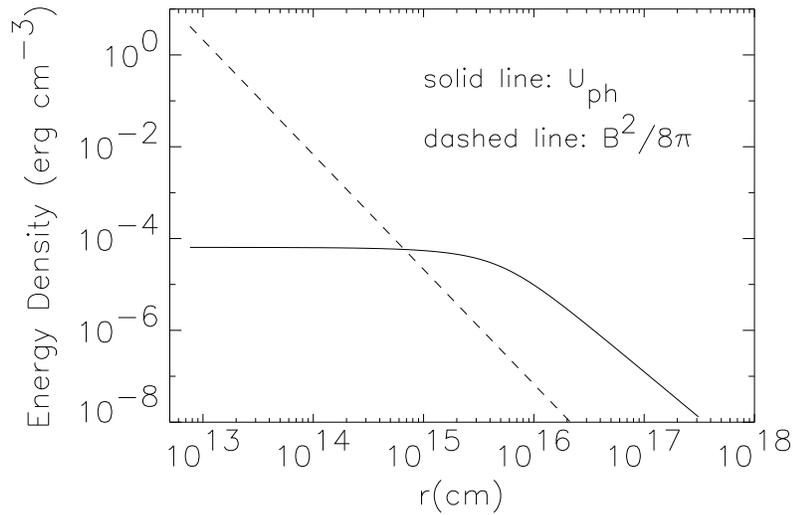}
\caption{Comparison between the energy density of the synchrotron emitted
photons $U_{ph}$ (solid line) and the energy density associated to the
magnetic field $B^2/8\pi$ (dashed line), obtained fixing $B_0=1$ $\mu$G and
ignoring the effect of advection (as in \cite{gon1,bert}).}
\label{fig:comp1}
\end{center}
\end{figure}

In this section, we reconsider the case of a spike with magnetic fields
as in \cite{gon1,bert}. We do this to show that these previous 
calculations have been carried out neglecting important pieces of physics,
that have now been introduced. We recall that the magnetic field assumed in 
\cite{gon1,bert} is much smaller than the {\it real} equipartition field 
defined in our Eq. (\ref{eq:magnetic}), although it was quoted in these 
previous papers as {\it the equipartition field}. This lower field could 
in principle be obtained assuming considerably lower accretion rate than  
is usually done.  

The results of \cite{gon1,bert} at high frequency are obtained ignoring  
the effect of advection and taking only synchrotron emission as the main 
channel for energy losses of the electrons with a magnetic field as in Eq. 
(\ref{eq:lowB}).
However, as we already discussed in previous sections, for such low fields, 
the effect of advection compared with synchrotron energy losses becomes very 
important, as follows from Eq. (\ref{eq:syn}): ignoring the contribution 
of advection results in the incorrect prediction for the luminosity of the 
generated radiations.

Another important point in the comparison with the results of \cite{gon1,bert}
is related to the role of Synchrotron Self-Compton scattering, which was also 
neglected in \cite{gon1,bert}. While SSC can indeed be neglected when 
using the correct equipartition magnetic field, as discussed in section
\ref{sec:equi} (see Fig. \ref{fig:Uph}), this is not true in the case of 
a lower magnetic field (cfr. Eq. (\ref{eq:lowB})) as the one used in 
\cite{gon1,bert}. In this latter case, SSC plays a crucial role. This can 
be easily shown by comparing the energy density of
synchrotron photons $U_{ph}(r)$ with the magnetic energy density 
$B_{low}^2/8\pi$. The result is shown in Fig. \ref{fig:comp1}, obtained
adopting $m_{\chi}=10^2$ GeV and $\langle \sigma v \rangle_{ann}=10^{-27}$ 
cm$^3$/s. From this figure it is evident that the effect of SSC is not 
negligible at distances larger than $\sim 10^3 R_g$ away from the GC. 
The effect of SSC is to lower the electron density outside about $10^2~R_g$ 
from the GC.

We confirm that the spectrum of synchrotron radiation is affected by strong 
synchrotron self-absorption at the low frequency end, as also pointed out in 
\cite{bert}.
This is an important point since in \cite{gon1} SSA was not included, which
led to an overestimate of the low frequency synchrotron luminosity.

\begin{figure}
\begin{center}
\includegraphics[width=0.7\textwidth]{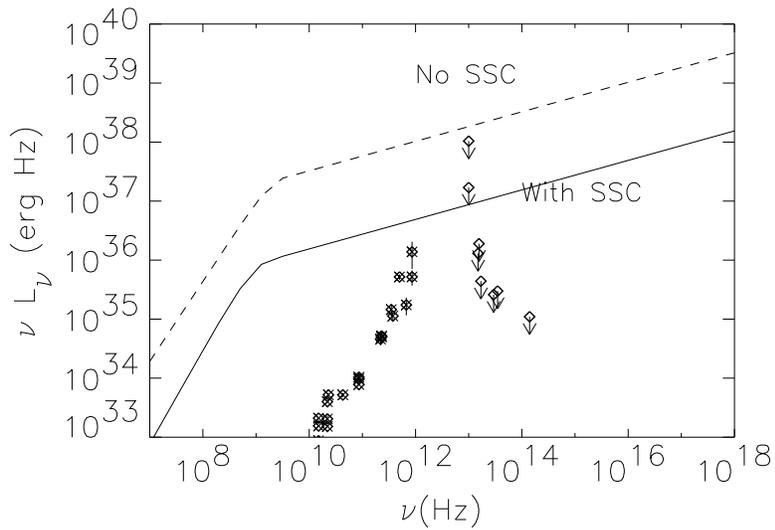}
\caption{Luminosity obtained ignoring the effect of advection and
taking into account the effect of SSC (continuos line) or ignoring it
(dashed line), compared with experimental data from Sgr A$^*$, in the case 
of the spike density profile with $B_0=1$ $\mu$G. }
\label{fig:comp2}
\end{center}
\end{figure}

In Fig. \ref{fig:comp2}  we neglect advection and compare observations 
with the emission as predicted including SSC, for $m_{\chi}=10^2$ GeV and 
$\langle \sigma v \rangle_{ann}=10^{-27}$cm$^3$/s. The effect of SSC is not 
enough to reconcile the predicted luminosity with observations. The spike 
hypothesis remains inconsistent with observations.

\section{Conclusions}
\label{sec:concl}

 In conclusion, we reaffirm that a spike density profile obtained from the 
adiabatic compression of a NFW profile in the gravitational field of the 
BH at the GC, using dark matter parameters typical of neutralinos, is ruled 
out by the radio and NIR observations from  SgA$^*$ while the NFW density 
profile remains a suitable hypothesis, not strongly constrained by available 
observations.

We reached this conclusion by a careful consideration of the accretion flow and
the loss processes in the transport equations for the neutralino generated
electrons and positrons as well as the radiative transfer. At the Galactic 
Center, the accretion flow onto the central BH sustains strong 
magnetic fields that induce synchrotron emission by electrons and positrons 
generated in neutralino annihilations during advection onto the BH. We found 
that the magnetic fields associated with the flow are significantly stronger
than previously adopted. With these equipartition fields, we derive the 
appropriate equilibrium distribution of electrons and positrons and the 
resulting radiation considering adiabatic compression in the accretion flow, 
inverse Compton scattering off synchrotron photons (synchrotron self-Compton 
scattering), and synchrotron self-absorption of the emitted radiation. We 
calculate the signal for a NFW dark matter halo profile and a NFW profile with 
a dark matter spike due to the central BH. We found that the annihilation 
of neutralino dark matter in the GC results in radio signals that overwhelm 
observations if the spike is present in the dark matter density profile. 
However, observed emissions from the GC are consistent with neutralinos 
following a NFW profile at the Galactic Center.

\section*{Acknowledgments}
We are grateful to Claudia Isola for a critical reading of the manuscript. 
The work of PB was partially funded through Cofin-2002/2003.
The work of AO was supported in part by the 
NSF through grant AST-0071235 and DOEgrant DE-FG0291-ER40606 
at the University of Chicago and at the Center for Cosmological 
Physics by grant NSF PHY-0114422.

\end{document}